\documentclass[aps,prc,superscriptaddress,twocolumn]{revtex4}
\usepackage{latexsym}
\usepackage{amsmath}  % 加载必要的数学包
\usepackage{amsfonts} % 加载 AMS（American Mathematical Society）提供的数学字体,常与amsmath一起使用
\usepackage{newtxtext,newtxmath} % 使用 Times New Roman 风格的字体 % 更改数学字体为 NewTXMath
\usepackage{bm} % 插入粗体数学符号
\usepackage{graphicx} % Required for inserting images
\usepackage{float} % 对浮动对象 图片和表格 的精确控制，[htbp]
\usepackage{subfigure}
\usepackage{epsf} % 插入eps格式图
\usepackage{epstopdf} % 确保 eps 转化为 pdf 格式
\usepackage{nameref} % 引用\label{} 编号对应的标题
\usepackage[colorlinks,
            citecolor=blue,
            anchorcolor=red,
            menucolor=red,
            linkcolor=red,
            filecolor=red,
            runcolor=red,
            urlcolor=blue,
            frenchlinks=true]{hyperref}
\usepackage{makecell} % 对表格格式的控制
\usepackage{array} % 提供 \extrarowheight 命令,\extrarowheight用来调整表格内容距离上行线的距离
\usepackage{xcolor} % color 的扩展包，包含了 color 包的所有功能，而且还提供了更多的扩展功能和灵活性
\usepackage{ulem} %用于提供文本的删除线（underline）和其他文本修饰功能

\begin{document}

\title{Effects of symmetry energy on the properties of hadron-quark mixed phase in hybrid stars}

\author{Min Ju}\email{jumin@upc.edu.cn}
\affiliation{School of Science, China University of Petroleum (East China), Qingdao 266580, China}

\author{Xuhao Wu}\email{wuhaoysu@ysu.edu.cn}
\affiliation{Key Laboratory for Microstructural Material Physics of Hebei Province, School of Science, Yanshan University, Qinhuangdao 066004, China}

\author{Hong Shen}\email{songtc@nankai.edu.cn}
\affiliation{School of Physics, Nankai University, Tianjin 300071, China}

\begin{abstract}
We study how the nuclear symmetry energy slope affects the properties of hadron-quark pasta phases and their sizes in massive hybrid stars ($> 2 M_{\odot}$).
We utilize the relativistic mean-field model with a density-dependent isovector coupling constant fitted by varying values of the symmetry energy slope $L$ to describe the hadronic matter, while the quark matter is described by a modified MIT bag model with vector interactions. We construct the hadron-quark phase transition using the energy minimization (EM) method and compare with the Gibbs construction (GC) in the hybrid star.
We find that the massive hybrid stars generally tend to have a hadron-quark pasta phase core rather than a pure quark core for most values of $L$, except in some special cases where they have a small pure quark core. 
Furthermore, we show that the mass and size of the pasta phase core are strongly influenced by the symmetry energy. The radius associated with the onset of pasta phases is also influenced by the symmetry energy. Additionally, we also evaluate the effects of the surface tension $\sigma$, the bag constant $B$, and the vector interaction $G_{V}$.
\end{abstract}

%生成文档的标题页，必须有
\maketitle  

%%%%%%%%%%%%%%%%%%%%%%%%%%%%%%%%%%%%%%%%%%%%%%%%%%%%%%%%%%%%%%%%%%%%%
\section{Introduction}
\label{sec:Intro}
Recent discoveries of pulsars with precisely measured masses close to 2 $M_{\odot}$, such as PSR J0348+0432 (with a mass of 2.01 $\pm$ 0.04 $M_{\odot}$ \cite{Antoniadis:2013pzd}) and PSR J0740+6620 (with a mass of 2.14$_{-0.09}^{+0.14}$ $M_{\odot}$ \cite{NANOGrav:2019jur}), have confirmed the existence of neutron stars with masses exceeding 2 $M_{\odot}$. 
In 2021, the mass measurement of PSR J0740+6620 was refined by the Neutron Star Interior Composition Explorer (NICER) to 2.08 $\pm$ 0.07 $M_{\odot}$ making it the highest reliably determined mass of any known neutron star. 
Concurrently, the star's radius was constrained within a range of 12.2 km $<$ $R$ $<$ 16.3 km \cite{Miller:2021qha} and 11.41 km $<$ $R$ $<$ 13.7 km \cite{Riley:2021pdl}. Further, a more recent study has provided tighter constraints on the radius, with an estimate of 11.61 km $<$ $R$ $<$ 13.77 km \cite{Salmi:2024aum}.  
The gravitational-wave event GW190814 was detected from the merger of binary coalescence involving a 23.2$_{-1.0}^{+1.1}$ $M_{\odot}$ black hole with a 2.59$_{-0.09}^{+0.08}$ $M_{\odot}$ compact object \cite{Fattoyev:2020cws}, where the secondary component could be interpreted as either the lightest black hole or the heaviest neutron star ever observed.

The central density of massive neutron stars can reach five to ten times the nuclear saturation density, providing a perfect environment for the existence of exotic forms of matter.
After decades of study, the composition of massive neutron stars continues to be an open question. Some studies indicate that hyperons are inevitable \cite{Djapo:2008au, Lopes:2020rqn, Tolos:2017lgv}. Hyperons may appear at 2 $\sim$ 3 times saturation density of nuclear matter, which could be suppressed by other degrees of freedom. Other possibilities include the presence of $\Delta$ resonances \cite{Ribes:2019kno, Motta:2019ywl} and kaon condensate \cite{Glendenning:1997ak, Sedrakian:2022ata}.
A more exotic scenario arises when the inner core of a neutron star undergoes a deconfinement phase transition, transforming into quark matter, while the outer layers continue to consist of hadronic matter \cite{Ju:2021nev, Ju:2021hoy, Rather:2020lsg}.
This represents a new branch of neutron stars known as hybrid stars, which typically have a quark core surrounded by hadronic matter \cite{Lopes:2021jpm, Li:2018ltg}. 
In Ref. \cite{Rather:2020lsg}, the authors demonstrated that a hadron-quark phase transition can reduce the maximum mass of hybrid stars by $\sim$ 0.5 $M_{\odot}$.
However, a model-independent analysis based on sound velocity suggests that quark cores are indeed expected inside massive neutron stars \cite{Annala:2019puf}.

An important quantity crucial to the equation of state (EoS) of neutron stars is the symmetry energy slope $L$ at saturation density ($n_0 \approx 0.16$ fm$^{-3}$). There is a consensus that there exists a positive correlation between the symmetry energy slope and the neutron-star radius \cite{Lattimer:2014sga, Cavagnoli:2011ft, Oertel:2016bki}. 
An even stronger correlation has been observed between the neutron skin thickness of heavy nuclei and the symmetry energy slope $L$ at subsaturation density ($n_b = 0.11$ fm$^{-3}$) \cite{Tews:2016jhi, Li:2008gp, Zhang:2013wna, Roca-Maza:2015eza}. 
These two correlations provide a unique bridge between atomic nuclei and neutron stars.
The constraints on the radius and the tidal deformability of a canonical 1.4 $M_{\odot}$ neutron star from astronomical observations favor a relatively smaller slope $L$, while the authors reported a value of $L$ = 106 $\pm$ 37 MeV \cite{Reed:2021nqk} using a specific class of relativistic energy density functionals based on the latest data of the neutron skin thickness of $^{208}$Pb from PREX-II.  
There are other calculations that infer $L$ from PREX-II, and different values have been obtained, such as $L=53^{+14}_{-15}$ MeV \cite{Essick:2021kjb}.
In Ref. \cite{Reinhard:2021utv}, the authors used some energy density functionals to study the parity violating asymmetry in $^{208}$Pb, directly measured by the PREX-II, to predict a smaller neutron skin thickness and inferred a smaller slope of the symmetry energy, $L$ =54 $\pm$ 8 MeV. 
Another experimental finding is the measurement of the spectra of charged pions that estimates $L$ to be in the range $42 < L< 117$ MeV \cite{SpiRIT:2021gtq}.
A revised analysis of PREX-II data, incorporating ground state properties of nuclei, astrophysical observations, and heavy-ion collision, indicates that the symmetry energy slope lies in the range 59 MeV $< L <$ 107 MeV \cite{Yue:2021yfx}.
Furthermore, a recent paper \cite{Tagami:2022spb} proposes two distinct ranges for 
$L$: one is in the range 0 $< L <$ 51 MeV according to the relatively thin neutron skin thickness of $^{48}$Ca reported by the CREX group \cite{CREX:2022kgg}, the other is in the range 76 MeV $< L < $ 165 MeV according to skin thickness of $^{208}$Pb from PREX-II \cite{PREX:2021umo}. 
There is significant uncertainty regarding the value of $L$ derived from CREX and PREX-II.
This is a big problem to be solved.
The above discussions have led to a common agreement that nuclear symmetry energy slope plays a crucial role in understanding various phenomena in nuclear physics and astrophysics. 

In this work, we aim to study hybrid stars based on the assumption that massive neutron stars should have a quark core, as suggested in Refs. \cite{Annala:2019puf, Lopes:2023dnx}.
We utilize relativistic mean-field model with a density-dependent isovector coupling constant fitted by varying values of the symmetry energy slope $L$  (referred to as RMFL model) to describe the hadronic matter, while the quark matter is described by a modified MIT bag model with vector interactions (referred to as vMIT bag model). The commonly used methods to describe hadron-quark phase transition involve the Gibbs construction (GC) and Maxwell construction (MC). 
In the GC \cite{Glendenning:1992vb}, the coexisting hadronic and quark phases are allowed to be charged separately, and the finite-size effects such as surface and Coulomb contributions are neglected.
When the surface tension of the hadron-quark interface is sufficiently large, the mixed phase is
close to those described by the MC, where local charge neutrality is imposed and the phase transition appears at the crossing of the hadronic EoS with the quark EoS in the pressure and chemical potential plane. The chemical potential at the intersection is also called critical chemical potential $\mu_0$, indicating the interface of the hadron-quark transition.
It is evident that GC and MC correspond to two limits of vanishing and large values of the surface tension, respectively, while both of them without considering the finite-size effects.
For a moderate surface tension, the hadron-quark mixed phase with some geometric structures, known as pasta phases, is expected to appear as a consequence of the competition between the surface and Coulomb energies.
An improved energy minimization (EM) method which is typically used for the description of the hadron-quark pasta phases incorporates the finite-size effects in a more consistent manner \cite{Wu:2018zoe}, where the equilibrium conditions for coexisting phases are derived by minimization of the total energy including surface and Coulomb contributions.
Further discussion can be found in our previous work \cite{ Ju:2021nev, Ju:2021hoy}.
It is worth mentioning that the pasta phases with a moderate surface tension have lower energy density compared to the MC, which means they are more stable. Although the energy density of pasta phases is slightly higher than that of GC, it is closer to the real phase transition process due to the consideration of the finite-size effects.
In this work, we employ the EM and GC methods to investigate the macroscopic properties of the hadron-quark mixed phase inside hybrid stars.

The main intention of this work is to study how the nuclear symmetry energy slope affects the properties of hadron-quark mixed phases and whether it is possible to have a pure quark core in massive hybrid stars. 
We use the RMFL model with NL3L parametrization for hadronic phase and vMIT model with an vector interaction for quark phase to ensure a massive hybrid star and it supports at least a hadron-quark mixed phase core.

This paper is organized in the following way.
The theoretical framework of RMFL model and vMIT bag model is presented in Section~\ref{sec:Theory}. 
We present numerical results and discuss how the symmetry energy slope affects the mass and the size percentage of the mixed phase core in Section~\ref{sec:results}.
Finally, the conclusions are provided in Section~\ref{sec:Conc}.
%%%%%%%%%%%%%%%%%%%%%%%%%%%%%%%%%%%%%%%%%%%%%%%%%%%%%%%%%%%%%%%%%%%%%
\section{formalism}
\label{sec:Theory}
We use RMFL model to describe the hadronic matter in which the isovector coupling constant $g_{\rho}$ is taken to be density dependent. 
The Lagrangian density for hadronic matter consisting of nucleons ($p$ and $n$) and leptons ($e$ and $\mu$) in the RMFL model has the following expression 
\begin{eqnarray}
\label{eq:LRMF} 
\mathcal{L_{\rm{RMFL}}}
& = & \sum_{i=n,p}\bar{\psi}_i \bigg\{i \gamma_{\mu}\partial^{\mu}- (M +g_{\sigma}\sigma) \notag\\
&&  -\gamma_{\mu} \left[g_{\omega}\omega^{\mu} 
   +\frac{g_{\rho}}{2}\tau_a\rho^{a\mu} \right] \bigg\}\psi_i  \notag \\
&& +\frac{1}{2}\partial_{\mu}\sigma\partial^{\mu}\sigma -\frac{1}{2}%
m^2_{\sigma}\sigma^2-\frac{1}{3}g_{2}\sigma^{3} -\frac{1}{4}g_{3}\sigma^{4} \notag \\
&& -\frac{1}{4}W_{\mu\nu}W^{\mu\nu} +\frac{1}{2}m^2_{\omega}\omega_{\mu}%
\omega^{\mu} \notag \\
&& -\frac{1}{4}R^a_{\mu\nu}R^{a\mu\nu} +\frac{1}{2}m^2_{\rho}\rho^a_{\mu}%
\rho^{a\mu} \notag \\
&& +\sum_{l=e,\mu}\bar{\psi}_{l}
\left( i\gamma_{\mu }\partial^{\mu }-m_{l}\right)\psi_l.
\end{eqnarray}  
$W^{\mu\nu}$ and $R^{\alpha\mu\nu}$ are the antisymmetric field tensors for $\omega^{\mu}$ and $\rho^{\alpha\mu}$, respectively.
The isovector coupling $g_{\rho}$ is taken to be density-dependent
\begin{equation}
    g_{\rho}(n_b) = g_{\rho}(n_0)\exp\left[-a_{\rho}\left(\frac{n_b}{n_0}-1\right)\right],
\end{equation}
where $n_0$ is the saturation density and $n_b$ is the baryon number density. $g_{\rho}(n_0)$ represents the value of $g_{\rho}$ at the saturation density $n_0$, which remains constant once the model parameters are fixed. 
So the symmetry energy slope $L$ can be tailored conveniently by adjusting $a_{\rho}$.
The Ref.~\cite{Wu:2021rxw} generates a series of parameter sets (referred to as NL3L) with different $L$ based on the NL3 parametrization. The parameters in NL3 model are given in Table \ref{tab:NL3}.
Adjusting \(L\) this way does not affect the saturation properties of nuclear matter. Therefore, the NL3L parameter sets have the same saturation properties as the original NL3 model, but with different values of \(L\). 
Table \ref{tab:NL3L} shows the parameter $a_\rho$ corresponding to the symmetry energy slope $L$ in the range of 30$-$118.5 MeV at saturation density. It should be noted that the case with $L$ = 118.5 MeV corresponds to the NL3 model.

The density dependence of $g_\rho$ contributes a rearrangement item
for nucleons,
\begin{flalign}
   \Sigma_r = \frac{1}{2} \sum_{i=n,p}\frac{\partial g_{\rho}(n_b)}{\partial n_{b} } n_i \tau_{3}\rho
    = -\frac{1}{2} a_\rho g_\rho(n_b)\frac{n_p - n_n}{n_0}\rho.
\end{flalign}
Correspondingly, the chemical potentials of nucleons can be expressed using the following equations
\begin{eqnarray}
\mu_p &=& \sqrt{(k_F^p)^2 + M^{*2}} + g_\omega \omega + \Sigma_r + \frac{g_\rho(n_b)}{2} \rho, \\
\mu_n &=& \sqrt{(k_F^n)^2 + M^{*2}} + g_\omega \omega + \Sigma_r - \frac{g_\rho(n_b)}{2} \rho,
\end{eqnarray}
where $M^* = M + g_\sigma \sigma $ is the effective nucleon mass, and $k_F^i$ is the Fermi momentum of species $i$, which is related to the vector density by $n_i = (k_F^i )^3/3\pi^2$. 
The EoS is then obtained in mean field approximation by calculating the components of the energy-momentum tensor. 
The total energy density and pressure for the hadron matter are given by
\begin{eqnarray}
\varepsilon_{\text{HP}} &=& \sum_{i=n,p} \frac{1}{\pi^2} \int_0^{k_F^i} \sqrt{k^2 + M^{*2} } k^2dk \notag\\
&& + \frac{1}{2} m_\sigma^2 \sigma^2 + \frac{1}{3} g_2 \sigma^3 + \frac{1}{4} g_3 \sigma^4 
   + \frac{1}{2} m_\omega^2 \omega^2  + \frac{1}{2} m_\rho^2 \rho^2  \notag\\
&& +\sum_{l=e,\mu}\frac{1}{\pi^2}\int_{0}^{k^{l}_{F}}dk k^2\sqrt{k^2+m_l^2},
\end{eqnarray}

\begin{eqnarray}
P_{\text{HP}} &=& \sum_{i=n,p} \frac{1}{3\pi^2} \int_0^{k_F^i} \frac{1}{\sqrt{k^2 + M^{*2}}} k^4 dk \notag\\
&& - \frac{1}{2} m_\sigma^2 \sigma^2 - \frac{1}{3} g_2 \sigma^3 - \frac{1}{4} g_3 \sigma^4 
+ \frac{1}{2} m_\omega^2 \omega^2 + \frac{1}{2} m_\rho^2 \rho^2  \notag\\
&&  + n_b \Sigma_r 
+ \sum_{l=e,\mu} \frac{1}{3\pi^2} \int_0^{k_F^l} \frac{1}{\sqrt{k^2 + m^{2}}} k^4 dk.
\end{eqnarray}
%%%%%%%%%%%%%%%%%%%%%%%%%%%%%%%%%%%%%%%%%%%%%%
\begin{table*}
\centering  %整个表格及其标题居中对齐
\hspace{0pt}
\setlength{\tabcolsep}{3.0mm}   % 调整列与列之间的间距
\addtolength{\extrarowheight}{5pt} % 调整表格文字距离上行线的距离，增加xpt
\caption{ Parameters in the NL3 model. The masses are given in MeV.}
\begin{tabular}{cccccccccc}
\hline\hline
Model & $M$ & $m_\sigma$ & $m_\omega$ & $m_\rho$ & $g_\sigma$ & $g_\omega$ & $g_\rho(n_0)$ & $g_2$ (fm$^{-1}$) & $g_3$ \\
\hline
NL3 & 939.000 & 508.194 & 782.501 & 763.000 & 10.217 & 12.868 & 8.948 & $-$10.431 & $-$28.885 \\
\hline\hline
\end{tabular}
\label{tab:NL3}
\end{table*}
%%%%%%%%%%%%%%%%%%%%%%%%%%%%%%%%%%%%%%%%%%%%%%%%

%%%%%%%%%%%%%%%%%%%%%%%%%%%%%%%%%%%%%%%%%%%%%%
\begin{table*}
\centering  %整个表格及其标题居中对齐
\setlength{\tabcolsep}{3.0mm}   % 调整列与列之间的间距
\addtolength{\extrarowheight}{5pt} % 调整距离上行线的距离，增加xpt
\caption{ Parameter $a_\rho$ generated from the NL3 model for different slope $L$ at saturation density $n_0$, while keeping other saturation properties unaffected~\cite{Wu:2021rxw}. 
The original NL3 model has $ L$ = 118.5 MeV.}
\begin{tabular}{ccccccccccc}
\hline\hline
$L$(MeV) & $30$ & $40$ & 50 & 60 & 70 & 80 & 90 & 100 & 110 & 118.5 \\
\hline
$a_\rho$ & 0.7537 & 0.6686 & 0.5835 & 0.4983 & 0.4132 & 0.3280 & 0.2429 & 0.1578 & 0.0726 & 0.0 \\
\hline\hline
\end{tabular}
\label{tab:NL3L}
\end{table*}
%%%%%%%%%%%%%%%%%%%%%%%%%%%%%%%%%%%%%%%%%%%%%%%%

%%%%%%%%%%%%%%%%%%%%%%%%%%%%%%%%%%%%%%%%%%%%%%%%%%%%%%%%%%%%%%%%%
For the quark phase at high densities, we consider it consisting of three flavor quarks ($u$, $d$, and $s$) and leptons ($e$ and $\mu$). We utilize the vMIT model to describe the quark phase, which has the following Lagrangian density
\begin{eqnarray}
\label{eq:LvMIT}
\mathcal{L_{\rm{vMIT}}} 
    &=& \sum_{q=u,d,s}\bigg[\bar{\psi}_{q}\left( i\gamma_{\mu}\partial^{\mu}
        -m_{q}-g_{qqV}\gamma_{\mu}V^{\mu}\right)\psi_{q} \nonumber  \\
    & & -B 
        +\frac{1}{2}m_{V}^{2}V_{\mu}V^{\mu}  \bigg]\Theta \nonumber  \\
    & & +\sum_{l=e,\mu}\bar{\psi_{l}}\left( i\gamma_{\mu}\partial^{\mu}
             -m_{l}\right)\psi_{l},
\end{eqnarray} 
where $m_q$ is the mass of the quark $q$ of flavor $u$, $d$, or $s$.
$B$ denotes the constant vacuum pressure and $\Theta$ is the Heaviside step function representing the confinement of quarks inside the bag.
The vector interaction is introduced via the exchange of a vector meson with the mass $m_V$
which is analogous to the $\omega$ meson in QHD \cite{Serot:1992ti}. 
The inclusion of vector channels in the MIT bag model is not new, as detailed in Ref. \cite{Lopes:2020btp}. 
With the help of mean field approximation, we obtained the equation of motion for the $V$ field,
\begin{eqnarray}
\label{eq:eqv0}
m_{V}^{2}V_{0} &=&  \sum_{q=u,d,s} g_{qqV}n_q ,
\end{eqnarray}
with $n_q$ being the number density of the quark flavor $q$. $g_{qqV}$ represents the interaction strength between the vector field with different quark flavors.
The quark chemical potential is then given by
\begin{eqnarray}
\label{eq:muq}
\mu_q=\sqrt{{k^{q}_{F}}^{2}+m_{q}^{2}}+g_{qqV}V_{0},
\end{eqnarray}
which is clearly enhanced by the vector potential.
The total energy density in quark matter is written as
\begin{eqnarray}
\varepsilon_{\rm{QP}} 
&=&   \sum_{q=u,d,s}
      \frac{N_c}{\pi^2}\int_{0}^{k^{q}_{F}}dk k^2\sqrt{k^2+m_q^2}
      +\frac{1}{2}m_V^2V_0^2     \nonumber \\
& &   +B +\sum_{l=e,\mu}\frac{1}{\pi^2}\int_{0}^{k^{l}_{F}}dk k^2\sqrt{k^2+m_l^2},
\label{eq:eqp}
\end{eqnarray}
where $N_c$ = 3 is the number of colors. 
%Moreover, the bag vacuum value is not independent of the vector field $V_0$, which ultimately depends on the strength of the coupling constant.
The pressure is obtained via the thermodynamic relation, $p = n\mu$ - $\varepsilon$, to guarantee thermodynamic consistency.
Here, we adopt the current quark masses $m_{u}=m_{d}=5.5$ MeV and $m_{s}=95$ MeV in our calculations following \cite{Lopes:2020btp}.

In the vMIT bag model, two important quantities, $X_V$ and $G_V$, are defined as
\begin{eqnarray}
    X_V = \frac{g_{ssV}}{g_{uuV}}, \qquad G_V = (\frac{g_{uuV}}{m_V})^2.
\end{eqnarray}
$X_V$ represents the relative strength of the vector field with the $s$ quark compared to the $u$ and $d$ quarks. 
There are two choices for $X_V$: one is $X_V = 0.4$ predicted by the symmetry group~\cite{Lopes:2020btp}, and the other is $X_V = 1.0$. The choice of $X_V = 1.0$ is under the assumption that $g_{qqV}$ is universal, implying the same value for all quark flavors, which is commonly used in \cite{Lopes:2019shs,Gomes:2018bpw,Menezes:2014aka,Ju:2021nev}.
Since this study primarily focuses on effects from different symmetry energy slope of the hadronic matter on the properties of mixed phase, using $X_V = 1.0$ is a better choice for maintaining the clarity and focus of the research. 
$G_V$ is related to the absolute strength of the vector field, and there are very few studies
trying to constrain its absolute value \cite{Restrepo:2014fna}. Most of the models just consider it as a free parameter \cite{Menezes:2014aka,Shao:2012tu}. 
The commonly used values of $G_V$  are in the range of 0 - 0.5 fm$^{2}$, while some larger values are also used in the study \cite{Ogihara:2019rds, Franzon:2016urz}. 
%\textcolor{red}{Our previous work \cite{ Ju:2021nev} demonstrated that a larger $G_V$ or $B$ leads to a higher onset density for the pasta phases, thereby supporting a more massive hybrid star.}
One way to constrain \( G_V \) is through the critical chemical potential $\mu_0$ in MC. However, so far, there are no experimental results that can indicate the critical chemical potential at \( T = 0 \), and our current understanding relies on effective models.
It was suggested in Ref. \cite{Lopes:2020dvs} that the critical chemical potential in MC is in the range of 1050 MeV $<$ $\mu_0$ $<$ 1400 MeV. The higher the value of $G_V$, the stronger the vector field interaction and the stiffer the EoS becomes, which leads to a higher $\mu_0$. 
Here we mainly adopt \( G_V = 0.2 \,\text{fm}^2 \), which results in the critical chemical potential $\mu_0 \sim$ 1300 MeV within MC based on NL3L parametrization. It is still lying in the range 1050 MeV $<$ $\mu_0$ $<$ 1400 MeV even in the extreme case for the NL3 model with $L$ = 118.5 MeV, which still supports the existence of relatively massive hybrid stars. 
On the other hand, the higher $G_V$ leads to a higher baryon density for the phase transition.
We use \( G_V = 0.1 \,\text{fm}^2 \) and \( G_V = 0.3 \,\text{fm}^2 \) for comparison.
As for the bag constant, it is well known that the bag constant could significantly affect the EoS of quark matter and consequently influence the hadron-quark phase transition. In this work, we aim to construct a hybrid star that has a quark core surrounded by hadronic matter. 
We primarily use $B^{1/4}$ = 180 MeV to investigate the effect of symmetry enery on the properties of mixed phase and also employ $B^{1/4}$ = 170 MeV for comparison.

%%%%%%%%%   pasta phases  %%%%%%%%%%%%%%%%%%%%%%%%%%%%%%
We compute the properties of hadron-quark
mixed phases by the EM method within the Wigner–Seitz approximation.
In the Wigner-Seitz approximation, the system is divided into a lot of equivalent and charge-neutral cells, where the coexisting hadronic and quark phases are separated by a sharp interface with finite surface tension. In general, for simplicity, the electron densities in the two coexisting phases are assumed to be spatially constant, and the charge screening effect is neglected. The details have been 
displayed in Ref. \cite{Ju:2021hoy}. 
Furthermore, in Ref. \cite{Ju:2021nev} we extended the EM method by allowing different electron densities in hadronic and quark phases to understand the transition from the GC to the MC.
A more realistic model for the structured mixed phase was proposed in Refs. \cite{Endo:2006ch,Tatsumi:2002dq}, in which the electric field was consistently treated within the Wigner-Seitz cell, leading to uneven distributions of charged particles in both hadronic and quark phases. However, treating the electric field in this manner is more time-consuming.
In the present work, we follow the method introduced in Ref. \cite{Ju:2021nev} to incorporate the charge screening effect in pasta phases for the sake of convenient calculation.

In the EM method, the total energy density of the mixed phase is expressed as
\begin{eqnarray}
\label{eq:fws}
\varepsilon_{\rm{MP}} &=& \chi \varepsilon_{\rm{QP}}
  + \left( 1 - \chi \right)\varepsilon_{\rm{HP}}
  + \varepsilon_{\rm{surf}} + \varepsilon_{\rm{Coul}} ,
\end{eqnarray}%
where $\chi=V_{\rm{QP}}/(V_{\rm{QP}}+V_{\rm{HP}})$ denotes the volume fraction of the quark phase.
The first two terms of Eq.~(\ref{eq:fws}) represent the bulk contributions,
while the last two terms come from the finite-size effects.
The surface and Coulomb energy densities are calculated from
\begin{eqnarray}
{\varepsilon}_{\rm{surf}}
&=& \frac{D \sigma \chi_{\rm{in}}}{r_D},
\label{eq:esurf} \\
{\varepsilon}_{\rm{Coul}}
&=& \frac{e^2}{2}\left(\delta n_c\right)^{2}r_D^{2} \chi_{\rm{in}}\Phi\left(\chi_{\rm{in}}\right),
\label{eq:ecoul}
\end{eqnarray}%
with%
\begin{eqnarray}
\label{eq:Du}
\Phi\left(\chi_{\rm{in}}\right)=\left\{
\begin{array}{ll}
\frac{1}{D+2}\left(\frac{2-D\chi_{\rm{in}}^{1-2/D}}{D-2}+\chi_{\rm{in}}\right),  & D=1,3, \\
\frac{\chi_{\rm{in}}-1-\ln{\chi_{\rm{in}}}}{D+2},  & D=2, \\
\end{array} \right.
\end{eqnarray}%
where $D=1,2,3$ denotes the geometric dimension of the cell, and $r_D$ represents
the size of the inner phase. $\chi_{\rm{in}}$ is the volume fraction of the inner phase,
i.e., $\chi_{\rm{in}}=\chi$ for droplet, rod, and slab configurations,
and $\chi_{\rm{in}}=1-\chi$ for tube and bubble configurations.
The charge density difference $\delta n_c$ between the hadronic and quark phases is defined as
\begin{equation}
    \delta n_c = n_c^{\text{HP}} - n_c^{\text{QP}},
\end{equation}
with
\begin{eqnarray}
    n_c^{\text{HP}} &=& n_p - n_e^{\text{HP}} - n_{\mu}^{\text{HP}},  \\
    n_c^{\text{QP}} &=& \frac{2}{3}n_u - \frac{1}{3}n_d - \frac{1}{3}n_s - n_e^{\text{QP}} - n_{\mu}^{\text{QP}}.
\end{eqnarray}
$\sigma$ denotes the surface tension at the hadron-quark interface, which plays a key role in determining the structure of the mixed phase.
The GC and MC, respectively, correspond to the two extreme cases of zero and large surface tension.
The hadron-quark pasta phases described in the EM method can be seen as an intermediate state between the GC and MC.
Currently, the value of \(\sigma\) is not well known, so it is usually taken as a free parameter. 
The study in Ref. \cite{Wu:2018zoe} shows that as $\sigma$ increases, the density range of hadron-quark mixed phase significantly shrinks and the number of pasta configurations reduces. 
A moderate surface tension could ensure that all the pasta phases appear as many as possible and make these pasta phases more stable than MC.
In this study, we examine three cases with surface tensions of $\sigma = 20$, $40$, and $60$ MeV fm$^{-2}$. Our primary focus is on $\sigma = 40$ MeV fm$^{-2}$, as this value is close to the prediction of the MIT bag model using the multiple reflection expansion method (MRE) \cite{Ju:2021hoy} and also ensures the appearance of all possible pasta phases.

At a given baryon density, the favored state is the one with the lowest energy among all possible configurations, which could be determined in the EM method by minimizing the total energy.
The equilibrium conditions for chemical potentials and pressures of the mixed phase can be found in Ref. \cite{Ju:2021nev}. 
We need to clarify that in this work, when discussing the EM method, we refer to structured mixed phase or pasta phases, which are typically called quark core in other studies \cite{Lopes:2021jpm, Lopes:2023dnx,Annala:2019puf,Li:2018ltg}. 

%%%%%%%%%%%%%%%%%%%%%%%%%%%%%%%%%%%%%%%%%%%%%%%%%%%%%%%%%%
\begin{figure*}[htbp]
\hspace{0pt}
\includegraphics[width=1.0\linewidth]{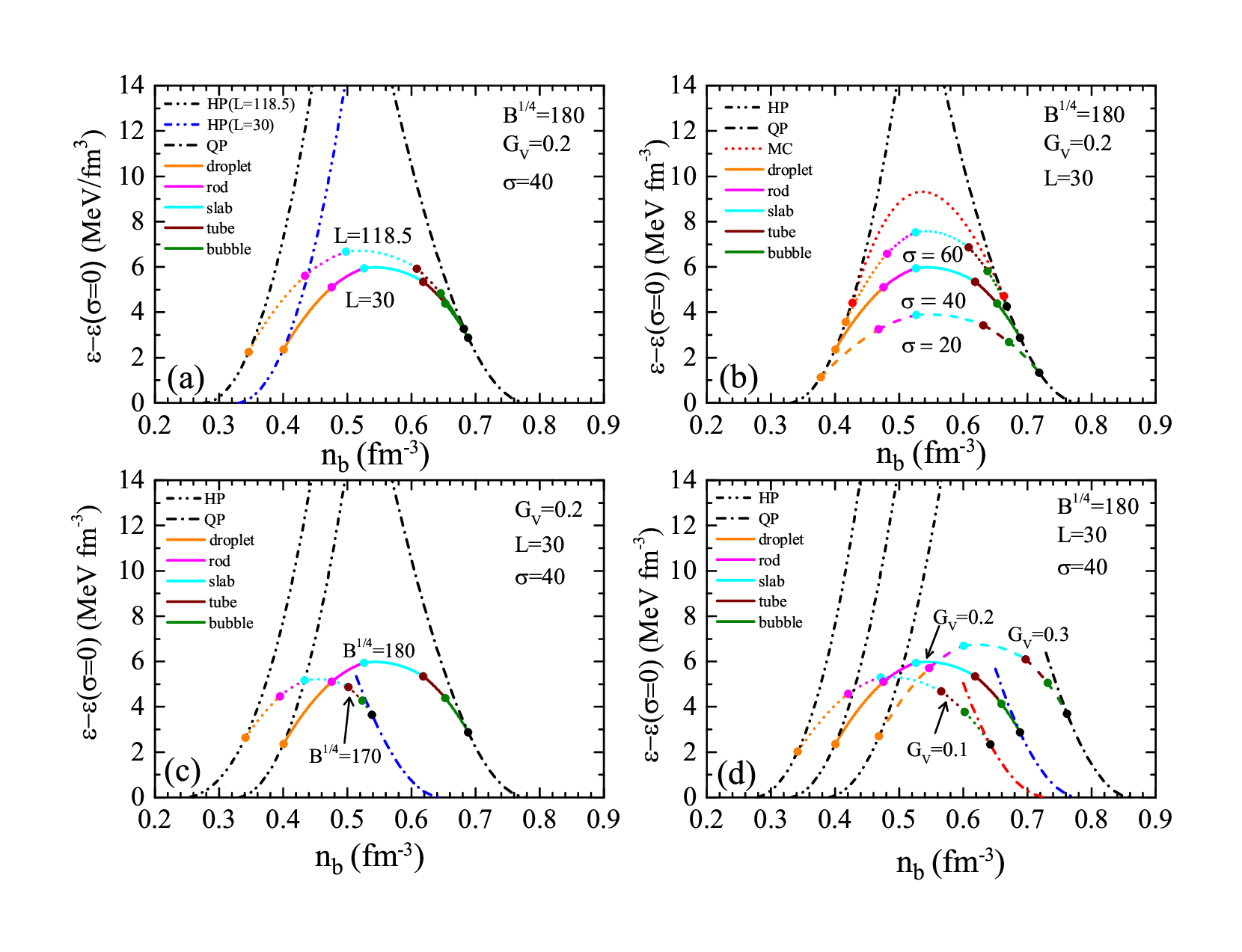}
\caption{ Energy densities of the pasta phases obtained using the EM method relative to those of the GC ($\sigma$ = 0) method. The filled circles indicate the transition points between different pasta phases.  }
\label{fig:1delta-ene}
\end{figure*}
%%%%%%%%%%%%%%%%%%%%%%%%%%%%%%

%%%%%%%%%%%%%%%%%%%%%%%%%%%%%%%%%%%%%%%%%%%%%%%
\begin{figure*}
\hspace{-90pt}
\centering
\begin{minipage}{0.5\textwidth}
    \includegraphics[bb=5 20 580 580, width=0.75\linewidth]{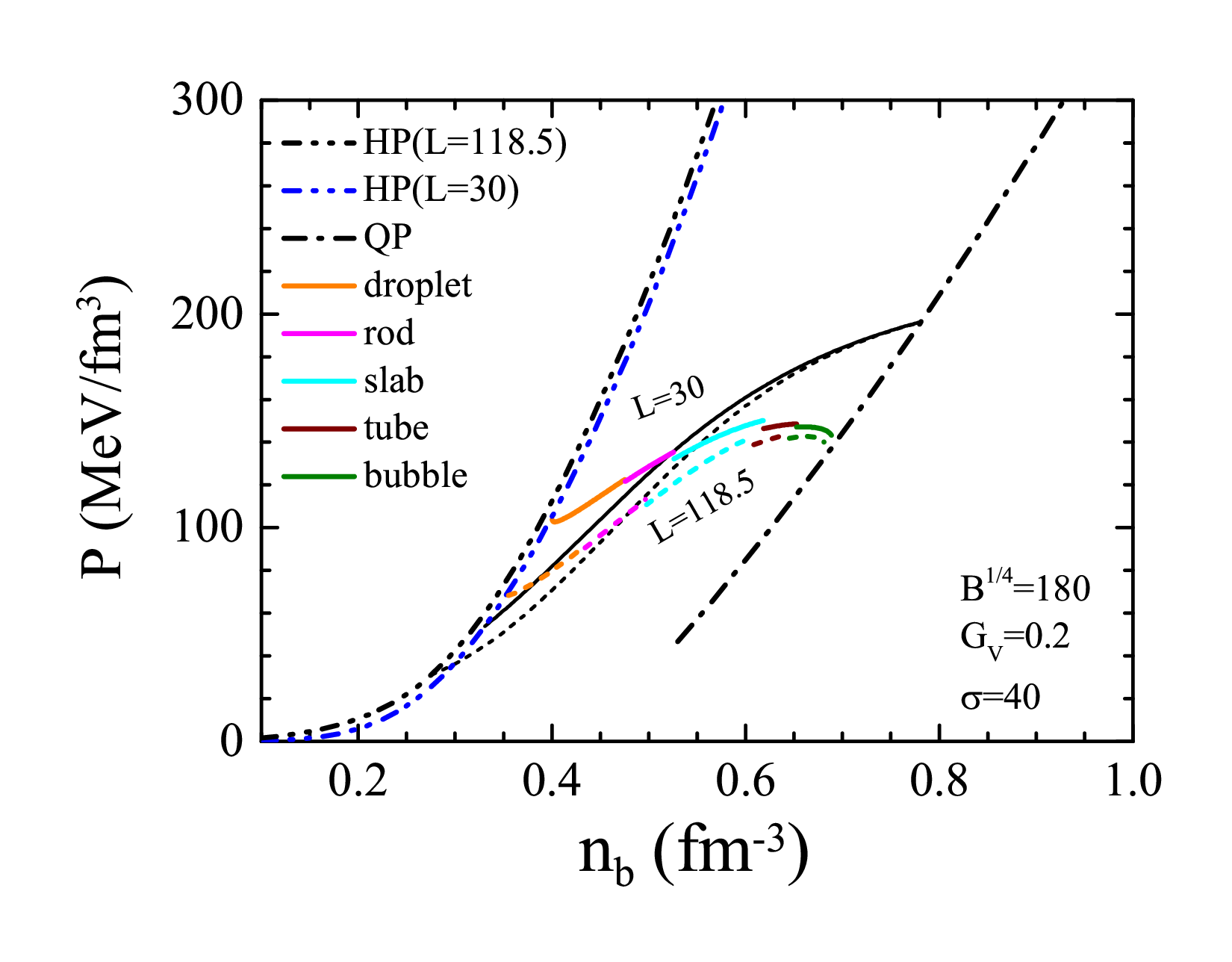}%
\end{minipage}
\begin{minipage}{0.5\textwidth}
    \includegraphics[bb=5 20 580 580, width=0.75\linewidth]{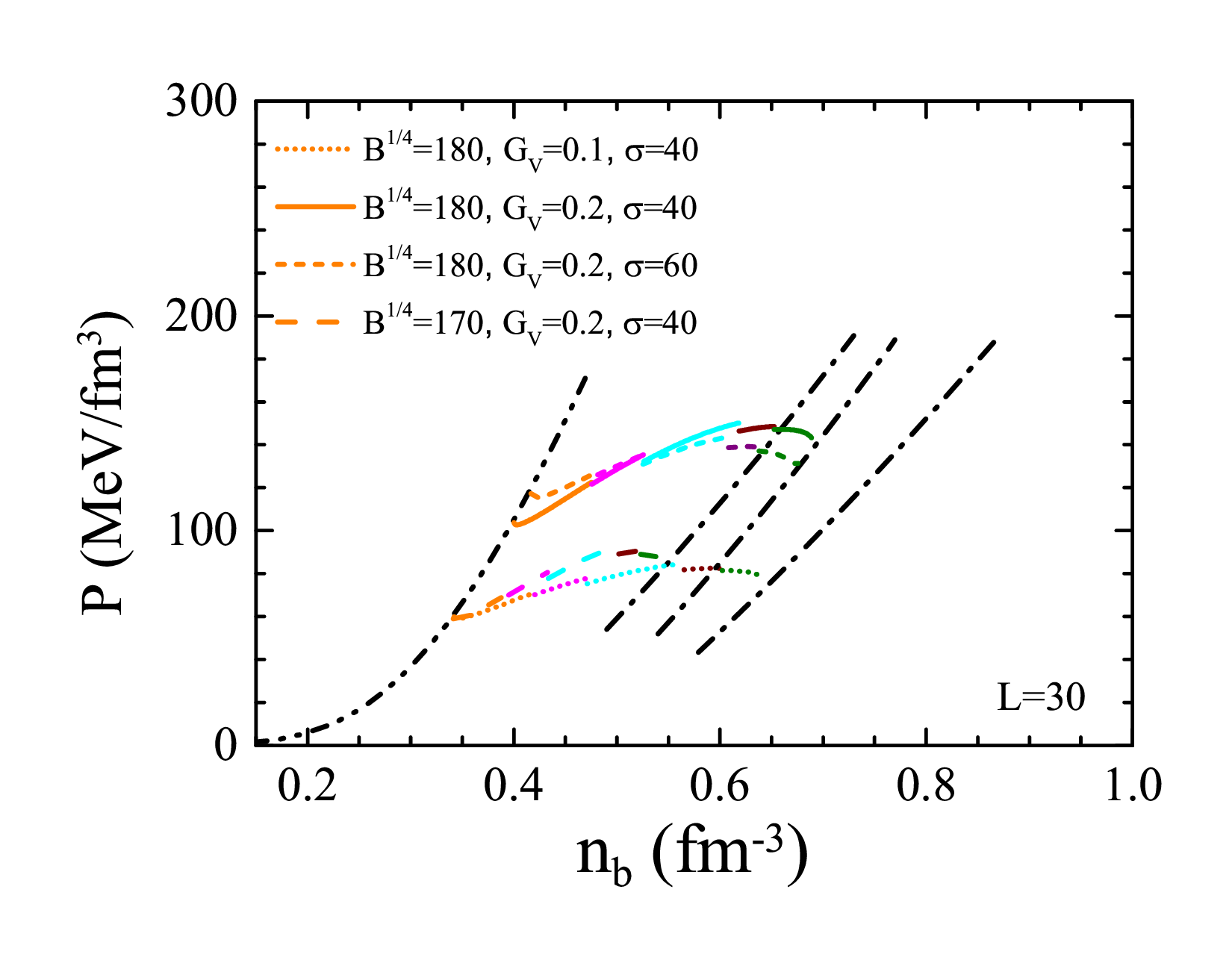}%
\end{minipage}
\caption{ Pressure as a function of baryon number density for hadronic, pasta, and quark phases. The solid (dashed) lines are the results of EM and GC methods with $L$ = 30 MeV ($L$ = 118.5 MeV) in the left panel. The results of EM method for different parameter choices are compared in the right panel.}
\label{fig:2P}
\end{figure*}
%%%%%%%%%%%%%%%%%%%%%%%%%%%%%%%%%%%%%%%%

%%%%%%%%%%%%%%%%%%%%%%%%%%%%%%%%%
\begin{table*}[htbp]
\caption{ Onset densities of the hadron-quark pasta phases and pure quark matter obtained using EM method with different model parameters.
\label{tab:GVBS1} }
\begin{center}
\setcellgapes{2.5pt} % 设置行间距为2pt, 与 \makegapedcells 搭配使用
\makegapedcells % 应用设置
\hspace*{0.0cm} % 调整这里的值来改变水平偏移量
\setlength{\tabcolsep}{2.5mm}{
\begin{tabular}{cccccccccc}
\hline\hline
\multicolumn{2}{c}{Model parameter}  &  Symmetry slope  &  Surface tension  
&  \multicolumn{6}{c}{Onset density (fm$^{-3}$)}     \\
\cline{1-2} \cline{5-10}
  $B^{1/4}$ (MeV)  &  $G_{V}$(fm$^{2}$)  &   $L$(MeV)  &  $\sigma$(MeV fm$^{-2}$)   
&Droplet   &Rod    &Slab   &Tube   &Bubble & Quark    \\
\hline
 180   & 0.1   & 30     &40  &0.342  &0.420 &0.471 &0.565 &0.602 &0.644  \\
 180   & 0.1   & 118.5  &40  &0.298  &0.390 &0.450 &0.557 &0.595 &0.636  \\
 \hline
 180   & 0.2   & 30     &40  &0.400  &0.476 &0.526 &0.619 &0.653 &0.689  \\
 180   & 0.2   & 118.5  &40  &0.346  &0.434 &0.500 &0.608 &0.645 &0.682  \\
 \hline
 170   & 0.2   & 30     &40  &0.341  &0.395 &0.433 &0.502 &0.523 &0.541  \\
 170   & 0.2   & 118.5  &40  &0.288  &0.354 &0.403 &0.489 &0.514 &0.532  \\
 \hline
 180   & 0.2   & 30     &20  &0.378  &0.468 &0.527 &0.631 &0.671 &0.719  \\
 180   & 0.2   & 30     &60  &0.417  &0.481 &0.525 &0.608 &0.638 &0.668  \\
\hline
\end{tabular} }
\end{center}
\end{table*}

%%%%%%%%%%%%%%%%%%%%%%%%%%%%%%

%%%%%%%%%%%%%%%%%%%%%%%%%%%%%%%%%
\begin{table*}[htbp]
\caption{The macroscopic properties of the maximum-mass hybrid star by using EM method with different model parameters.
\label{tab:GvBS2} }
\begin{center}
\setcellgapes{2.5pt} % 设置行间距为2.5pt, 与 \makegapedcells 搭配使用
\makegapedcells % 应用设置
\hspace*{0.0cm} % 调整这里的值来改变水平偏移量
\setlength{\tabcolsep}{2mm}{
\begin{tabular}{cccccccccc}
\hline\hline
\multicolumn{2}{c}{Model parameter}  &  Symmetry slope  &  Surface tension  
&   $M_{\text{max}}$ & $M_{\text{MP}}$ & $\Delta M_{\text{MP}}$ & $R_{\text{max}}$ & $R_{\text{MP}}$ & $\Delta R_{\text{MP}}$      \\
\cline{1-2}
$B^{1/4}$ (MeV)  &  $G_{V}$(fm$^{2}$)  &   $L$(MeV)  &  $\sigma$(MeV fm$^{-2}$)   
&($M_{\odot}$) & ($M_{\odot}$) & ($M_{\text{MP}}$/$M_{\text{max}}$) & (km) & (km) &  ($R_{\text{MP}}$/$R_{\text{max}}$)    \\
\hline
 180   & 0.1   & 30     &40  &1.917  &0.197 &10.3\% &13.60 &5.004 &36.8\%  \\
 180   & 0.1   & 118.5  &40  &1.813  &0.391 &21.6\% &14.42 &6.341 &44.0\%  \\
 \hline
 180   & 0.2   & 30     &40  &2.263  &0.956 &49.2\% &13.11 &8.288 &63.2\%  \\
 180   & 0.2   & 118.5  &40  &2.178  &1.294 &59.4\% &13.18 &9.365 &71.1\%  \\
 \hline
 170   & 0.2   & 30     &40  &2.014  &1.005 &49.9\% &12.18 &7.538 &61.9\%  \\
 170   & 0.2   & 118.5  &40  &1.945  &1.373 &70.6\% &11.91 &8.387 &70.4\%  \\
 \hline
 180   & 0.2   & 30     &20  &2.307  &0.442 &19.2\% &13.53  &6.219  &46.0\%  \\
 180   & 0.2   & 30     &60  &2.355  &0.114 &4.8\% &13.71  &3.869  &28.2\%  \\
\hline
\end{tabular} }
\end{center}
\end{table*}

%%%%%%%%%%%%%%%%%%%%%%%%%%%%%%

%%%%%%%%%%%%%%%%%%%%%%%%%%%%%%%%%%%%%%
\begin{figure*}
\hspace{-30pt}
\includegraphics[ width=1.05\linewidth]{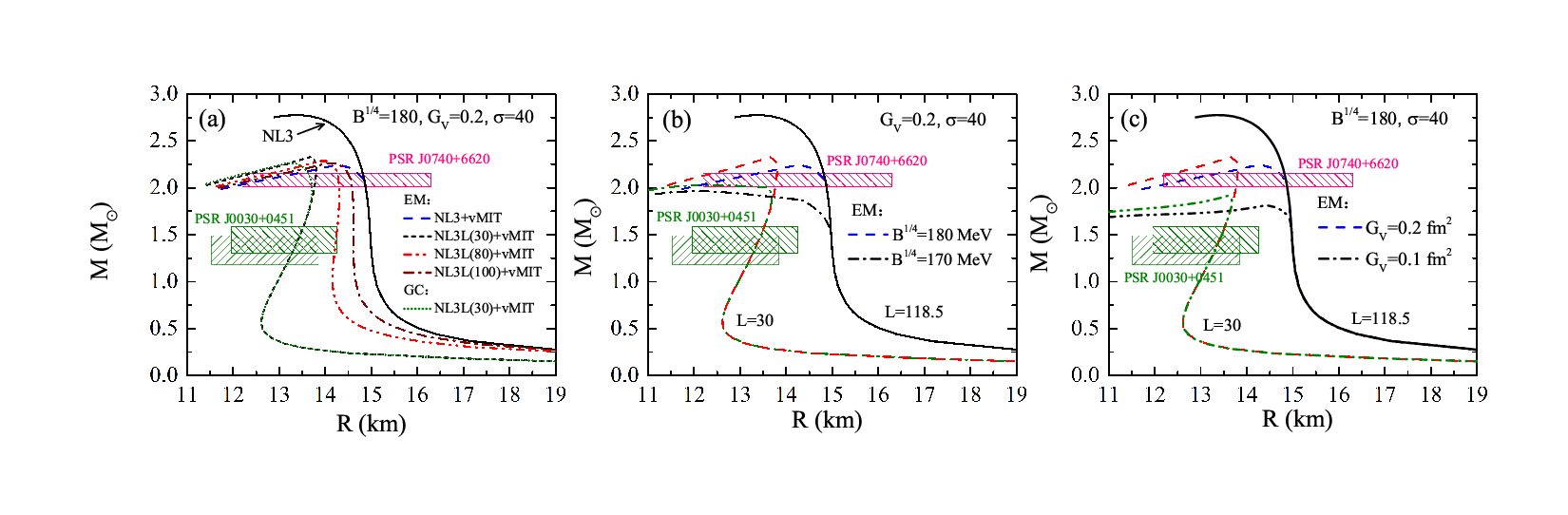}%
\caption{ Mass-radius relations for hybrid stars obtained using EM method with different parameter choices. In panel (a), the values in parentheses indicate the corresponding $L$ for the hadronic phase. 
The black solid line represents the results of pure hadronic matter obtained by NL3 model.
%The green dotted line is the results from GC method with $L$ = 30 MeV for comparison. 
The green and pink hatched areas correspond to the constraints from NICER for PSR J0030+0451 \cite{Miller:2019cac,Riley:2019yda} and PSR J0740+6620 \cite{Miller:2021qha}, respectively. 
}
\label{fig:3mr}
\end{figure*}
%%%%%%%%%%%%%%%%%%%%%%%%%%%%%%%%%%%%%%%%

%%%%%%%%%%%%%%%%%%%%%%%%%%%%%%%%%%%%%%%%%%%%%%%
\begin{figure*}
\hspace{-90pt}
\centering
\begin{minipage}{0.52\textwidth}
    \includegraphics[bb=5 20 580 580, width=0.75\linewidth]{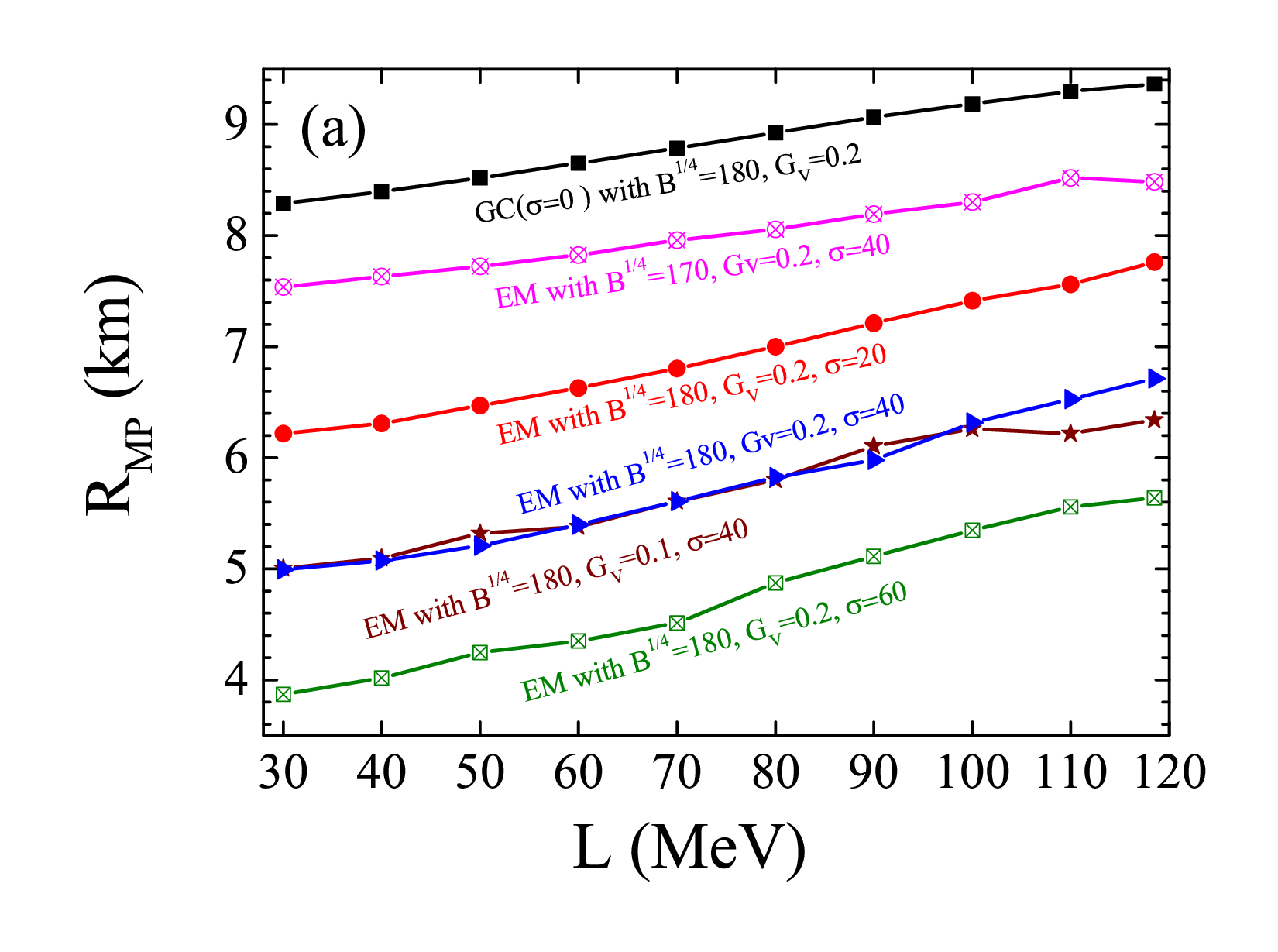}%
\end{minipage}
\hspace{-20pt} % 调整这里的数值来缩小间距
\begin{minipage}{0.52\textwidth}
    \includegraphics[bb=5 20 580 580, width=0.75\linewidth]{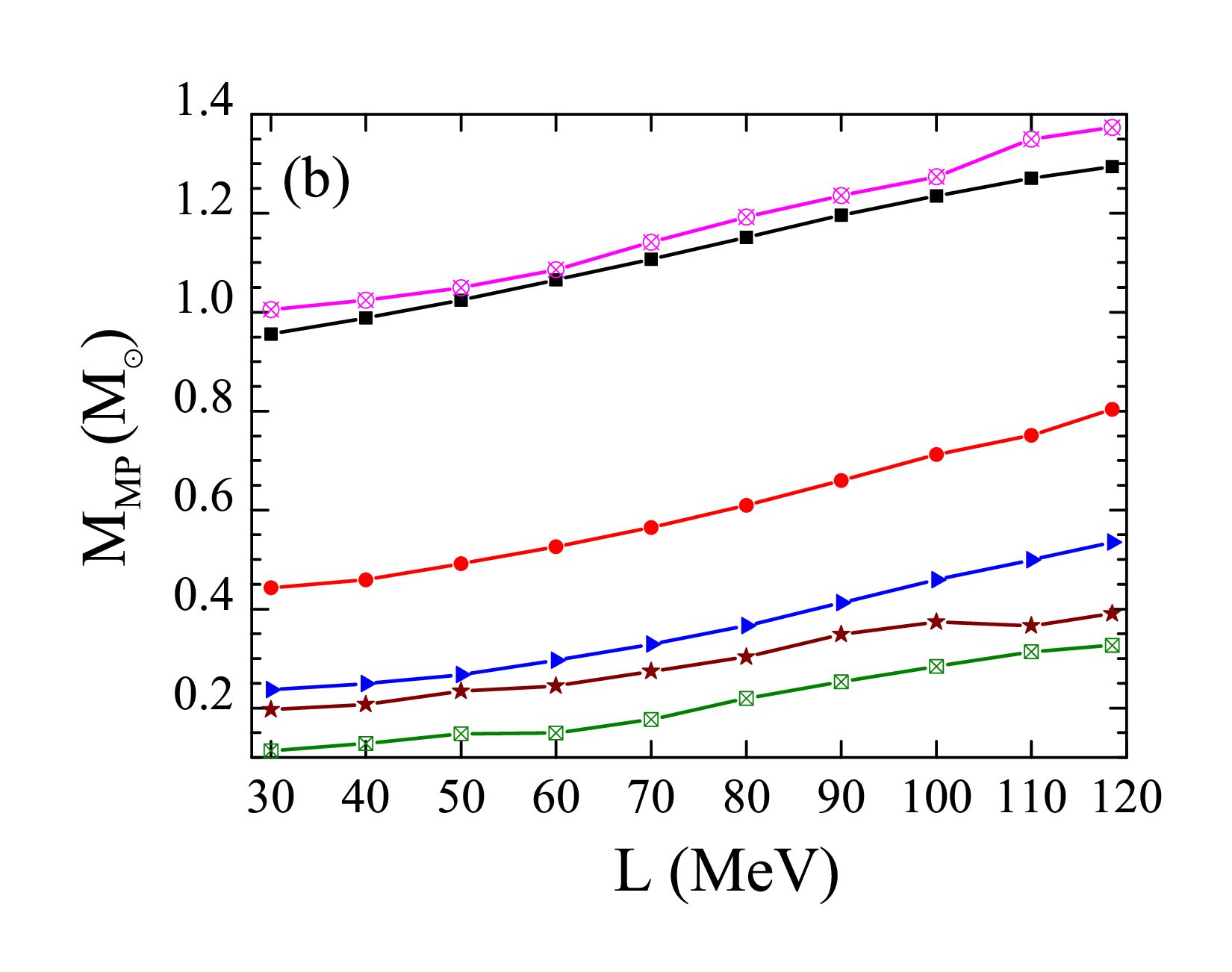}%
\end{minipage}
\caption{ The radius (a) and mass (b) of the structured mixed phase using EM method in the maximum-mass hybrid star as a function of the symmetry energy slope $L$. The black line is the result of GC method. }
\label{fig:4RMmp}
\end{figure*}
%%%%%%%%%%%%%%%%%%%%%%%%%%%%%%%%%%%%%%%%
%%%%%%%%%%%%%%%%%%%%%%%%%%%%%%%%%%%%%%%%%%
\begin{figure*}
\hspace{0pt}
\includegraphics[width=1.0\linewidth]{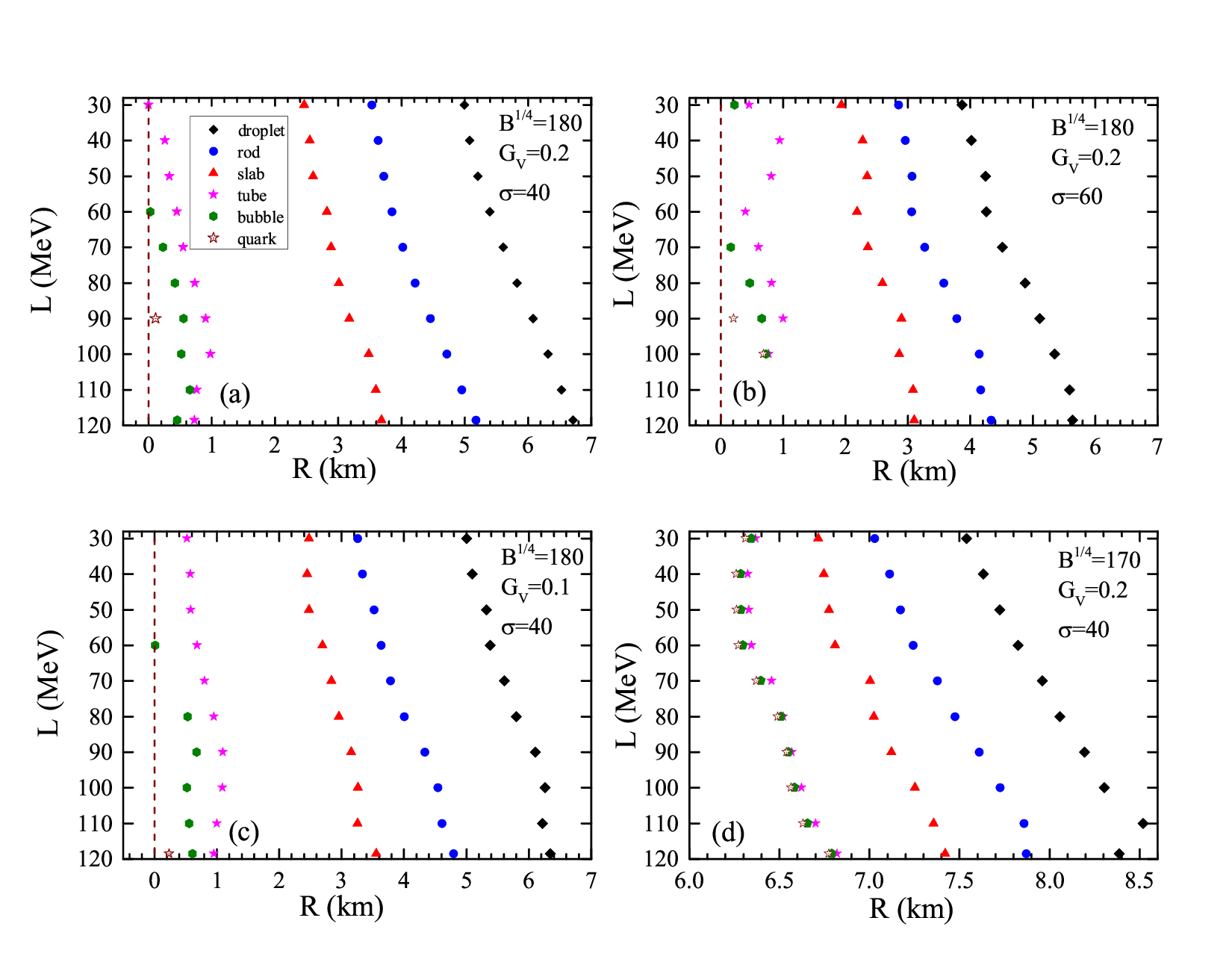}%
\caption{ The initial appearance radius of the pasta phases and pure quark phase relative to the center of the maximum-mass hybrid star. The results are obtained by using the EM method. The vertical dashed line is plotted to help identify the center of hybrid star. }
\label{fig:5Rpasta}
\end{figure*}
%%%%%%%%%%%%%%%%%%%%%%%%%%%%%%%%%%%%%%%%
%%%%%%%%%%%%%%%%%%%%%%%%%%%%%%%%%%%%%%%%%%%%%%%%%%%%%%%%%%%%%%%%%%%%%%%%%%%%%%%%
\section{Results and discussion}
\label{sec:results}
We utilize the RMFL model with the NL3L parameterization to describe the hadronic matter, while
the quark matter is described by the vMIT bag model. The hadron-quark pasta phases are computed by the EM method and the properties of hybrid stars are calculated by using the EoS with pasta phases.
In Fig. \ref{fig:1delta-ene}, we show the energy densities of the hadronic phase, pasta phases and the quark phase relative to those of the GC ($\sigma$ = 0) method with different parameter choices.
The onset density of the mixed phase from the GC method is indicated by the starting points of the dash-dot-dot lines of HP, whereas the orange filled circles represent the onset density of the pasta phases as determined by the EM method.
One can see that the finite-size effects considered in EM method, increase the energy density and delay the appearance of the pasta phases compared to the onset of mixed phase determined by the GC method. 
In Fig. \ref{fig:1delta-ene} (a), the dash-dot-dot lines represent the EoS of hadron phase with $L$ = 30 MeV (blue line) and $L$ = 118.5 MeV (black line), while the dash-dot line is the EoS of quark phase based on the vMIT model with $G_V$ = 0.2 fm$^{2}$.
The pasta phases appear when its energy density becomes lower than that of hadronic phase. It is shown that the pasta configuration changes from droplet to rod, slab, tube, and bubble with increasing baryon density $n_b$.
Meanwhile, the energy difference between the EM and GC methods increases with the symmetry energy slope $L$. On the other hand, the difference between $L$ = 30 MeV and $ L$ = 118.5 MeV decreases as the density increases, disappearing when the pure quark matter becomes the most stable state. 
This implies the symmetry energy slope mainly affects the onset densities of pasta phases, while it hardly influences the appearance density of pure quark phase, i.e., the pasta phases begin to appear at 0.346 fm$^{-3}$(and end at 0.682 fm$^{-3}$) for $L$ = 118.5 MeV and they begin at 0.400 fm$^{-3}$(and end at 0.689 fm$^{-3}$) for $L$ = 30 MeV, as depicted in Table \ref{tab:GVBS1}.
That is anticipated because in pasta phases, the fraction of hadronic matter decreases with increasing baryon density, which results in the impacts from the hadronic matter getting weaker, and drops to zero when the pure quark phase becomes the most stable state. 
Fig. \ref{fig:1delta-ene} (b) shows the results for different surface tensions $\sigma$, with the red dashed line representing the results of MC. As $\sigma$ increases, the energy density of the pasta phases also increases, leading to a higher onset density of the pasta phases, as depicted in Table \ref{tab:GVBS1}. Notably, when $\sigma = 60$ MeV fm$^{-2}$, both the onset and end densities of the pasta phases closely match those obtained from the MC. This suggests that $\sigma = 60$ MeV fm$^{-2}$ is approaching the upper limit for surface tension. 
The effects of bag constant $B$ and vector interaction $G_V$ are illustrated in Fig. \ref{fig:1delta-ene} (c) and (d), as well as in Table \ref{tab:GVBS1}. It can be observed that smaller values of $B$ or $G_V$ have similar effects on the pasta phases. The pasta phases appear and end earlier for smaller values of $B$ and $G_V$, which is consistent with the results in Ref.~\cite{Ju:2021nev}. Specially, for $B^{1/4}$ = 170 MeV, the quark phase appears much earlier, e.g., when $L$ = 118.5 MeV, it begins at 0.532 fm$^{-3}$.
It is worth mentioning that for the case with $G_V$ = 0.3 fm$^{2}$, the critical potential $\mu_0$ in MC exceeds 1500 MeV. Therefore, we focus mainly on the cases with $G_V$ = 0.1 fm$^{2}$ and $G_V$ = 0.2 fm$^{2}$.

We show in Fig. \ref{fig:2P} the pressure $P$ as a function of baryon density for hadronic, pasta and quark phases. The results from GC method with different $L$ are shown for comparison in the left panel. The pressure of pasta phases with $L$ = 118.5 MeV is lower than the case for $L$ = 30 MeV, which can account for the earlier appearance of pasta phases for larger $L$. Moreover, the density range of pasta phases obtained by the EM method is noticeably narrower than that obtained by the GC method, and the EM method seems to generate a soft phase transition compared with GC method. 
From the right panel in Fig. \ref{fig:2P}, we find that both $B$ and $G_V$ can affect the properties of quark matter. As a result, earlier appearances of pasta phases are observed in the cases with $B^{1/4}$ = 170 MeV (compare to $B^{1/4}$ = 180 MeV) and $G_V$ = 0.1 fm$^2$  (compare to $G_V$ = 0.2 fm$^2$).
A larger surface tension $\sigma$ has a small effect on the pressure and slightly reduces the density range of pasta phases.

In Fig. \ref{fig:3mr}, we show the the mass-radius relations for hybrid stars. 
We also discuss a couple of observational constraints inferred from the NICER x-ray telescope. The first, and maybe the most important one, is the PSR J0740+6620, whose mass lies in the range of $M$ = 2.08 $\pm$ 0.07 $M_{\odot}$. The radius has two main ranges: one is 11.41 km $<$ $R$ $<$ 13.70 km \cite{Riley:2021pdl}, the other one is 12.2 km $<$ $R$ $<$ 16.3 km \cite{Miller:2021qha}. Here, we use the less restrictive approach, which is the latter one, as the constraint. 
%The allowed region $R_{1.4} \leq $ 13.6 km, inferred from GW170817, is also shown. From Fig. \ref{fig:5mr}, one can see that this constraint is not satisfied when $L \geq$ 100 MeV. 
From Fig. \ref{fig:3mr} (a), we observe that when we consider the hadron-quark phase transition, the increase of $L$ results in a decrease in the maximum mass of hybrid stars by approximately 0.1 $M_{\odot}$ between $L$ = 30 MeV and 118.5 MeV, as depicted in Table \ref{tab:GvBS2}. This is consistent with the results from Ref. \cite{Lopes:2023dnx}, where the Quantum Hadrodynamics model is used to describe the hadron phase and the vMIT model is utilized for quark phase but with larger values of $G_V$. 
In Ref. \cite{Lopes:2023dnx}, the authors used the MC method for hadron-quark phase transition without considering the structured mixed phase.
The original NL3 model predicts a maximum mass of 2.77 $M_{\odot}$. With a hadron-quark phase transition, the maximum mass of the NL3L model is reduced to around 2.26 $M_{\odot}$. 
Although the radius increases as $L$ increases, they still satisfy the constraint from PSR J0740+6620 \cite{Miller:2021qha} and EoSs with smaller symmetry energy ($L \leq $ $80$ MeV) satisfy the constraints from PSR J0030+0451 \cite{Miller:2019cac}.
From Fig. \ref{fig:3mr} (b) and Fig. \ref{fig:3mr} (c), we can observe that smaller $B$ or $G_V$ result in a lower mass hybrid star, as also depicted in Table \ref{tab:GvBS2}. For $B^{1/4}$ = 170 MeV, only the case $L$ = 30 MeV can reach 2 $M_{\odot}$. When $L$ =118.5 MeV, it leads to the highest percentage of radius and mass for the hadron-quark pasta phase core, where both the mass and the radius of the pasta phase core are approximately 70\% of the total mass and radius, respectively. 
From Table \ref{tab:GVBS1} and \ref{tab:GvBS2}, one can find that a larger $\sigma$ delays the onset density of pasta phases but leads to an earlier end density. As a result, a larger $\sigma$ leads to a smaller radius for the pasta phases and 
consequently increases slightly the maximum mass of hybrid stars.

Now we estimate the size and mass of the mixed phase core in the maximum-mass hybrid star.
In Fig. \ref{fig:4RMmp}, we show the radius and mass of mixed phase core obtained by the EM and GC methods as a function of the symmetry energy slope $L$. One can find that both the radius and mass of the mixed phase increase with the slope $L$. Additionally, the results from the GC method are clearly larger than those from the EM method within the range 30 MeV $\leq$ $L$ $\leq$ 118.5 MeV,  for example, with a difference of approximately 4 km for the radius and approximately 0.8 $M_{\odot}$ for the mass with $B^{1/4}$ = 180 MeV, $G_V$ = 0.2 fm$^2$, and $\sigma = 40$ MeV fm$^{-2}$. 
The smaller mass and radius within the EM method are due to the higher onset density and narrower density range of the hadron-quark mixed phase compared to that in the GC method. 
Specifically, a smaller $B$ results in a larger pasta phase core, with its mass being slightly higher than that obtained from the GC method.
With decreasing $\sigma$, the radius and mass of the pasta phase core approach those obtained from the GC method. This can be attributed to the reduced surface tension, which decreases the contribution from finite-size effects.
The size of the pasta phase core with $G_V$ = 0.1 fm$^{2}$ is comparable to that with $G_V$ = 0.2 fm$^{2}$, but it has a slightly lower pasta phase mass.

%%%%%%%%%%%%%%%%%%%%%%%%%%%%%%%%%%%%%%%%%%%%%%%%%%%%%%%%
% GC 方法得到的混杂相性质统计表格
\begin{table*}
\setcellgapes{3pt} % 设置表格及表格间的额外空间, 表格内容仍保持居中，based on the package makecell
\makegapedcells % 应用设置
\hspace*{0cm} % 调整这里的值来改变水平偏移量
\setlength{\tabcolsep}{2.5mm}   % 调整列与列之间的间距, based on none package
\caption{ Onset densities of the mixed phase $n_1$ and pure quark phase $n_2$, and the macroscopic properties of the maximum-mass hybrid star. The results are obtained using the GC method with $B^{1/4}$ = 180 MeV and $G_V$ = 0.2 fm$^{2}$. $R_{\text{max}}$ corresponds to the radius of the maximum-mass hybrid star, while $n_{c}$ is the central density. $R_{\text{MP}}$ and $M_{\text{MP}}$ denote the radius and the mass of the mixed phase, respectively. $\Delta M_{\text{MP}}$ ($\Delta R_{\text{MP}}$) represents the mass (radius) percentage of the mixed phase within the maximum-mass hybrid star. }
\centering
      \begin{tabular}{ccccccccccc}
      \hline\hline
          Model & $L$ & $n_{1}$ & $n_{2}$ & $n_{c}$ 
          & $M_{\text{max}}$ & $M_{\text{MP}}$ & $\Delta M_{\text{MP}}$ & $R_{\text{max}}$ & $R_{\text{MP}}$ & $\Delta R_{\text{MP}}$   \\
              & (MeV) & (fm$^{-3}$) & (fm$^{-3}$) & (fm$^{-3}$) & ($M_{\odot}$) & ($M_{\odot}$) & ($M_{\text{MP}}$/$M_{\text{max}}$) & (km) & (km) &  ($R_{\text{MP}}$/$R_{\text{max}}$)   \\
       \hline
        & 30   & 0.331 & 0.781 & 0.678 & 2.263 & 0.956 & 42.2\% & 13.11 & 8.288 & 63.2\% \\
        & 40   & 0.328 & 0.781 & 0.678 & 2.257 & 0.988 & 43.8\% & 13.12 & 8.399 & 64.0\%\\
        & 50   & 0.323 & 0.781 & 0.678 & 2.250 & 1.024 & 45.5\% & 13.13 & 8.519 & 64.9\%\\
        & 60   & 0.318 & 0.781 & 0.678 & 2.242 & 1.065 & 47.5\% & 13.14 & 8.655 & 65.9\%\\
 NL3L   & 70   & 0.312 & 0.781 & 0.679 & 2.234 & 1.109 & 49.6\% & 13.15 & 8.787 & 66.8\% \\
        & 80   & 0.305 & 0.781 & 0.679 & 2.224 & 1.151 & 51.8\% & 13.16 & 8.929 & 67.8\% \\
        & 90   & 0.298 & 0.781 & 0.680 & 2.214 & 1.196 & 54.0\% & 13.17 & 9.069 & 68.9\% \\
        & 100  & 0.290 & 0.781 & 0.681 & 2.202 & 1.235 & 56.1\% & 13.18 & 9.189 & 69.7\% \\
        & 110  & 0.282 & 0.781 & 0.682 & 2.189 & 1.271 & 58.1\% & 13.19 & 9.299 & 70.5\% \\
        \hline
  NL3  & 118.5 & 0.275 & 0.781 & 0.684 & 2.178 & 1.294 & 59.4\% & 13.18 & 9.365 & 71.1\% \\
           \hline\hline
      \end{tabular}
      \label{tab:GC}
\end{table*}
%%%%%%%%%%%%%%%%%%%%%%%%%%%%%%%%%%%%%%%%%%%%%%%%%%%%%%%%
%%%%%%%%%%%%%%%%%%%%%%%%%%%%%%
% EM 方法得到的pasta性质统计表格
\begin{table*}
\setcellgapes{3pt} % 设置表格及表格间的额外空间, 表格内容仍保持居中，based on the package makecell
\makegapedcells % 应用设置
\hspace*{0cm} % 改变水平偏移量
\setlength{\tabcolsep}{4mm}   % 调整列与列之间的间距，based on none package
\caption{ Onset densities of the hadron-quark pasta phases and pure quark phase within the EM method. The results are obtained with $B^{1/4}$ = 180 MeV, $G_V$ = 0.2 fm$^{2}$, and $\sigma = 40$ MeV fm$^{-2}$.   }
      \begin{tabular}{ccccccccc}
      \hline\hline
      Model & $L$ &\multicolumn{6}{c}{Onset density (fm$^{-3}$)} & $n_{c}$  \\
      \cline{3-8} % 用于在表格中绘制部分水平线
      & (MeV) &  Droplet &  Rod & Slab & Tube & Bubble  &  Quark  & (fm$^{-3}$) \\
          \hline
           & 30    & 0.400 & 0.476 & 0.526 & 0.619 & 0.653 & 0.689 & 0.617 \\
           & 40    & 0.399 & 0.474 & 0.525 & 0.618 & 0.653 & 0.689 & 0.632 \\
           & 50    & 0.396 & 0.473 & 0.524 & 0.618 & 0.652 & 0.689 & 0.640 \\
           & 60    & 0.392 & 0.470 & 0.523 & 0.617 & 0.652 & 0.689 & 0.652 \\
    NL3L   & 70    & 0.388 & 0.467 & 0.521 & 0.617 & 0.652 & 0.688 & 0.665 \\
           & 80    & 0.381 & 0.463 & 0.519 & 0.616 & 0.651 & 0.688 & 0.680 \\
           & 90    & 0.373 & 0.457 & 0.515 & 0.615 & 0.650 & 0.687 & 0.689 \\
           & 100   & 0.364 & 0.450 & 0.510 & 0.613 & 0.649 & 0.685 & 0.670 \\
           & 110   & 0.354 & 0.442 & 0.504 & 0.611 & 0.647 & 0.684 & 0.666 \\
           \hline
      NL3  & 118.5 & 0.346 & 0.434 & 0.500 & 0.608 & 0.645 & 0.682 & 0.677 \\
           \hline\hline
      \end{tabular}
      \label{tab:EM-ni}
\end{table*}
%%%%%%%%%%%%%%%%%%%%%%%%%%%%%%%%%%%%%%%%%%%%%%%%%%%%%%%%
%%%%%%%%%%%%%%%%%%%%%%%%%%%%%%
% EM 方法得到的pasta性质统计表格
\begin{table*}
\setcellgapes{3pt} % 设置表格及表格间的额外空间, 表格内容仍保持居中，based on the package makecell
\makegapedcells % 应用设置
\hspace*{0cm} % 调整这里的值来改变水平偏移量
\setlength{\tabcolsep}{4mm}   % 调整列与列之间的间距,based on none package 
\caption{ The macroscopic properties of the maximum-mass hybrid star by using the EM method. The results are obtained with $B^{1/4}$ = 180 MeV, $G_V$ = 0.2 fm$^{2}$, and $\sigma = 40$ MeV fm$^{-2}$.  }
      \begin{tabular}{cccccccc}
      \hline\hline
     Model & $L$ & $M_{\text{max}}$ & $M_{\text{MP}}$ & $\Delta M_{\text{MP}}$ & $R_{\text{max}}$ & $R_{\text{MP}}$ & $\Delta R_{\text{MP}}$ \\
           & (MeV) & ($M_{\odot}$) & ($M_{\odot}$) & ($M_{\text{MP}}$/$M_{\text{max}}$) & (km) & (km) &  ($R_{\text{MP}}$/$R_{\text{max}}$) \\
     \hline
           & 30    & 2.334 & 0.237 & 10.2\% & 13.50 & 4.99  & 37.0\% \\
           & 40    & 2.327 & 0.249 & 10.7\% & 13.52 & 5.07  & 37.5\%  \\
           & 50    & 2.320 & 0.267 & 11.5\% & 13.53 & 5.21  & 38.5\%   \\
           & 60    & 2.311 & 0.296 & 12.8\% & 13.55 & 5.40  & 40.0\% \\
    NL3L   & 70    & 2.303 & 0.329 & 14.3\% & 13.57 & 5.61  & 41.3\%  \\
           & 80    & 2.293 & 0.366 & 16.0\% & 13.60 & 5.82  & 42.8\%  \\
           & 90    & 2.284 & 0.413 & 18.1\% & 13.64 & 5.98  & 43.8\%  \\
           & 100   & 2.269 & 0.459 & 20.2\% & 13.68 & 6.32  & 46.2\%  \\
           & 110   & 2.255 & 0.499 & 22.1\% & 13.73 & 6.53  & 47.6\%  \\
           \hline
      NL3  & 118.5 & 2.243 & 0.535 & 23.9\% & 13.78 & 6.71 & 48.7\% \\
           \hline\hline
      \end{tabular}
      \label{tab:EM-mr}
\end{table*}
%%%%%%%%%%%%%%%%%%%%%%%%%%%%%%%%%%%%%%%%%%%

%%%%%%%%%%%%%%%%%%%%%%%%%%%%%%%%%%%%%%
\begin{figure*}
\hspace{-10pt}
\includegraphics[width=1.02\linewidth]{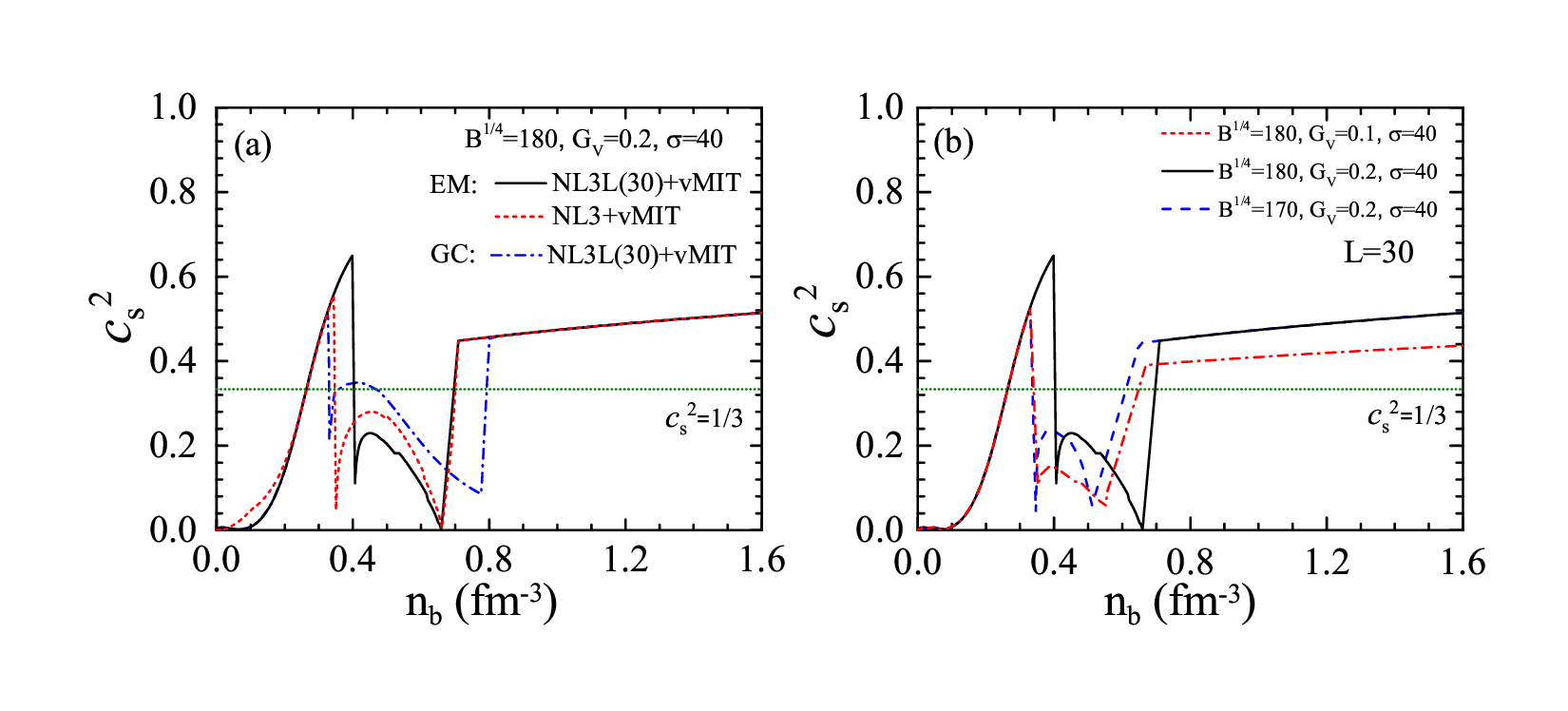}%
\caption{ The squared speed of sound as a function of baryon number density by using EM and GC methods for different parameter choices in left panel. The results of EM method for different parameter choices are compared in right panel.
The dotted line $c_s^2 =$ 1/3 indicates the conformal limit. 
}
\label{fig:6vs2}
\end{figure*}
%%%%%%%%%%%%%%%%%%%%%%%%%%%%%%%%%%%%%%%%

%%%%%%%%%%%%%%%%%%%%%%%%%%%%%%%%%%%%%%
\begin{figure*}
\hspace{-10pt}
\includegraphics[width=1.02\linewidth]{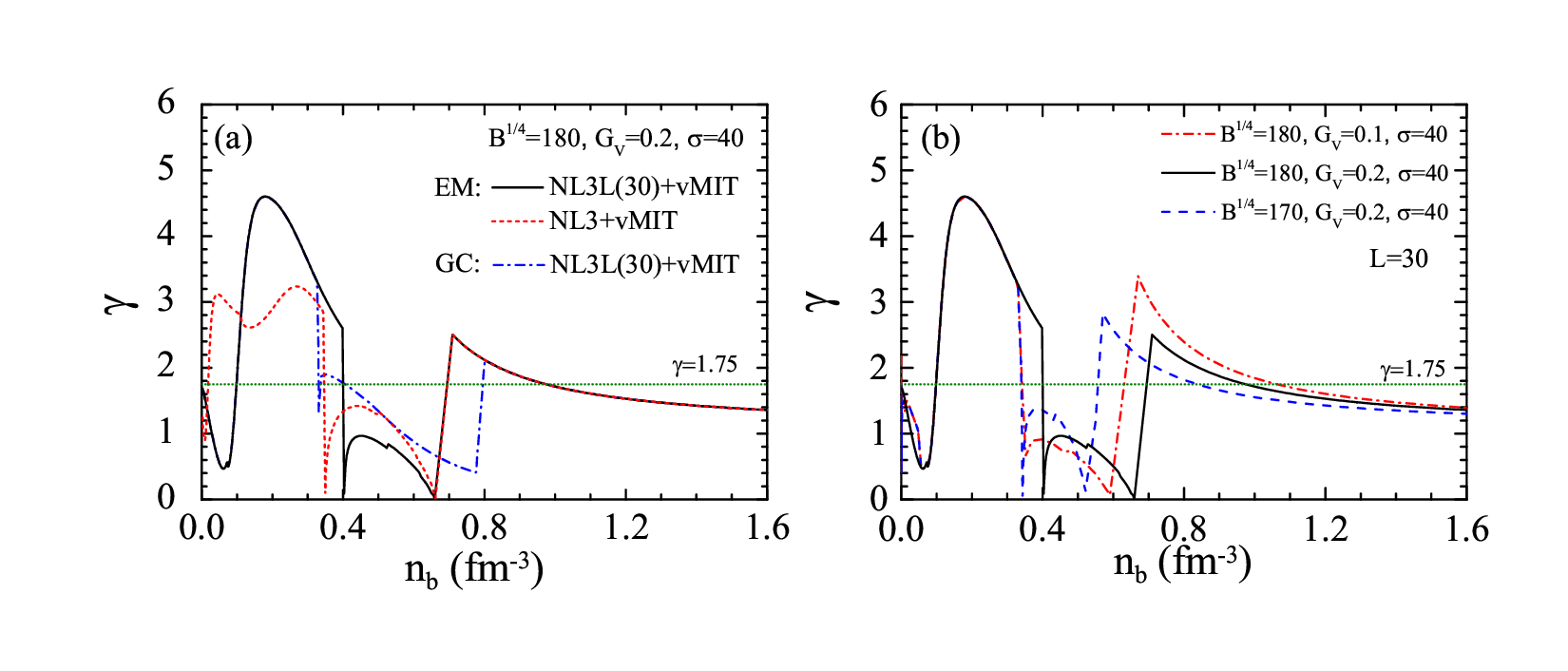}%
\caption{ The polytropic index as a function of baryon number density by using EM and GC methods in left panel. The results of EM method for different parameter choices are compared in right panel. 
%The dotted line $\gamma =$ 1.75 corresponds to the conformal limit. 
}
\label{fig:7gamma}
\end{figure*}
%%%%%%%%%%%%%%%%%%%%%%%%%%%%%%%%%%%%%%%%

We present in Table \ref{tab:GC} the onset densities of the mixed phase $n_1$ and pure quark phase $n_2$, the central density $n_c$ and the macroscopic properties of the maximum-mass hybrid star by using GC method.
As we can see, only the onset density of the mixed phase is shifted to a lower range by increasing $L$, while the end density of the mixed phase is hardly influenced, corresponding to a wider density range. The central density of the maximum-mass hybrid star is less than the onset density of pure quark phase, indicating that the maximum-mass hybrid star prefers to have a mixed phase core rather than a pure quark core under GC.
Furthermore, we can observe that with increasing $L$ from 30 MeV to 118.5 MeV, the mass of the mixed phase core increase from approximately 42\% to approximately 59\% of the total mass of the maximum-mass hybrid star, while the corresponding radius increase from approximately 63\% to approximately 71\%.  

In Table \ref{tab:EM-ni}, we present the onset densities of pasta phases (droplet, rod, slab, tube, bubble), pure quark phase and the central density of the maximum-mass hybrid star obtained by the EM method. One can see that as $L$ increases, the whole pasta phases are shifted to lower baryon density, together with a wider density range. That is because that $L$ has a slighter effect on the onset density of pure quark phase than on the onset of pasta phases. It is worth mentioning that for EM method, the central density does not monotonically increase with $L$ like the results of GC method. Instead, there is a inflection point at $L \sim$ 90 MeV, which causes the fact that only for $L$ = 90 MeV does the maximum-mass hybrid star have a small pure quark core $\sim$ 0.11 km.
The macroscopic properties of the maximum-mass hybrid star obtained by using EM method are presented in Table \ref{tab:EM-mr}.
The variation trends of the mass and the radius of pasta phase core are the same as those of GC method, however the percentage is visibly smaller. For instance, for $L$ = 60 MeV, the mass of pasta phase core is 0.296 $M_{\odot}$ (12.8\%) by the EM method, while it rises to 1.065 $M_{\odot}$ (47.5\%) by the GC method, correspondingly, the radius varies from 5.4 km (40.0\%) to 8.7 km (65.9\%).
Nevertheless, in all cases, a massive hybrid star would have a pasta phase core with mass above 0.24 $M_{\odot}$ and a radius above 5 km. Meanwhile, the maximum mass of hybrid stars decreases with the increase of $L$, but the corresponding radius is slightly increased, same tendency is observed by the GC method.

To further study the detailed properties of pasta phases in the maximum-mass hybrid star, we present the initial appearance radius of the pasta phases (droplet, rod, slab, tube, bubble) relative to the center of hybrid stars for different $L$ in Fig. \ref{fig:5Rpasta}.
From Fig. \ref{fig:5Rpasta} (a), one can find that for droplet, rod and slab phases, the appearance radius increases with the symmetry energy slope $L$. However, for the tube and bubble phases, there is a turning point at $L$ = 90 MeV, which is caused by the instability of the central density when $L$ $>$ 90 MeV.
Moreover, it is easy to find that for $L$ $<$ 60 MeV, the hybrid stars have a pasta phase core containing droplet, rod, slab and tube phases. The bubble phase would appear in the core of hybrid stars when $L$ $\geq$ 60 MeV, but no pure quark matter found in hybrid stars, except for $L$ = 90 MeV, where the maximum-mass hybrid star would have a small pure quark core, as denoted by the hollow star shape.
From Fig. \ref{fig:5Rpasta} (b) with $\sigma = 60$ MeV fm$^{-2}$, we observe that a pure quark core exists only for $L$ = 90 MeV and $L$ = 100 MeV. In contrast, for $\sigma = 20$ MeV fm$^{-2}$, only a pasta phase core is present, while the pure quark core cannot be formed.
In Fig. \ref{fig:5Rpasta} (c) with $G_V$=0.1 fm$^2$, it is shown to have a pasta phase core with a radius around 6.0 km. However, when $L$ = 118.5 MeV, it seems to have a small pure quark core.
In Fig. \ref{fig:5Rpasta} (d) with $B^{1/4}$ = 170 MeV, all the cases for 30 MeV $\leq$ $L$ $\leq$ 118.5 MeV support a pure quark core with a radius exceeding 6.0 km, while the radii of pasta phases are significantly larger than those shown in Fig. \ref{fig:5Rpasta} (a) with $B^{1/4}$ =180 MeV. 

The squared speed of sound, $c_s^2 = \partial P / \partial \varepsilon$, is related to the stiffness of the EoS, and can give us an important insight into the internal composition of neutron stars.
In general, for a pure nucleonic neutron star, the squared speed of sound grows monotonically. However, the introduction of new degrees of freedom can lead to nontrivial behavior, causing the squared speed of sound to exhibit maxima and minima.
For the heaviest reliably observed neutron stars with mass $M \sim 2 M_\odot$, the presence of quark matter is found to be linked to the behavior of the squared speed of sound $c_s^2$ in strongly interacting matter.
A constant $c_s^2$ = 1/3 corresponds to exactly conformal matter, such as strange quark matter at high densities \cite{Bedaque:2014sqa}.
It seem to have an agreement that if the conformal bound $c_s^2$ $\leq$ 1/3 is not strongly violated, massive hybrid stars are predicted to have sizable quark-matter cores \cite{Annala:2019puf}.
We illustrate the squared speed of sound as a function of baryon density for hybrid stars with a hadron-quark mixed phase core for different slope $L$ by using EM method and GC methods in Fig. \ref{fig:6vs2} (a). The results of EM method for different $B$ and $G_V$ are displayed in Fig. \ref{fig:6vs2} (b).
One can see that the squared speed of sound increases with the baryon density in the hadronic phase. 
There is a sudden reduction in the squared speed of sound after the appearance of the pasta phases or mixed phase, as a result of more degrees of freedom. The squared speed of sound remains below 1/3 throughout the pasta phases and drops to zero at the end of these phases. 
The results from GC method are higher than these of EM method, which is due to the consideration of finite-size effects in the EM method.
The effects from $L$ on $c_s^2$ of the hadronic phase are significant only at densities below 0.2 fm$^{-3}$, and a larger slope $L$ would increase the squared speed of sound of pasta phases.
At higher densities where the quark phase appears, $c_s^2$ is around 0.5 which is higher than 1/3, that is because we considered the vector interaction among quarks.
It is not a serious concern because, based on the models used in this work, the central density of hybrid stars does not reach such high densities.
Another physical quantity, which is also considered to be a good approximate criterion for the evidence of quark matter in neutron stars, is the polytropic index $\gamma = d(\ln{P}) / d(\ln{\varepsilon})$.
The polytropic index has the value $\gamma$ = 1 in conformal matter, while the hadronic models generically predict $\gamma \approx$ 2.5 around and above saturation density \cite{Kurkela:2009gj}.
Given that $\gamma$ = 1.75 is both the average between its perturbative Quantum Chromodynamics and Chiral Effective Theory limits, and very close to the minimal value the quantity obtains in viable hadronic models \cite{Annala:2019puf}. $\gamma$ = 1.75 is used as an approximate criterion for indicating the appearance of quark matter.
As one can see in Fig. \ref{fig:7gamma}, the polytropic index $\gamma$ of the pasta phases is below 1.75, while there is a sharp rise after the pure quark matter shows up, then it stabilized to $\gamma$  $\approx$ 1.4. 
The variation of the polytropic index during the mixed phase is relatively smooth in the GC method compared to that in EM method, because the GC method assumes a uniform mixed phase without structures.

%%%%%%%%%%%%%%%%%%%%%%%%%%%%%%%%%%%%%%%%%%%%%%%%%%%%%%%%%%%%%%%%%%%%%%%%%%%%%%%%%%%%%%%%%%%
\section{Conclusions}
\label{sec:Conc}
Motivated by the observations of massive neutron stars and the possibility of a hadron-quark phase transition in their cores, we investigated how changes in the slope of the symmetry energy affect the mass and size of the mixed phase core in hybrid stars, including possible geometric structures. 
For the hadronic phase, we utilized a modified version of the NL3 model which includes a density-dependent $g_{\rho}$ to adjust the symmetry energy slope $L$ without altering other saturation properties. For the quark phase, we employed a modified MIT bag model with vector interactions. The EM method, in which the equilibrium conditions for coexisting phases are derived by minimizing the total energy including surface and Coulomb contributions, was used to describe the structured hadron-quark mixed phase. The results from the GC method, which can be simply understood as representing a uniform hadron-quark mixed phase, were also presented for comparison.

We analyzed the structured hadron-quark mixed phase using the EM method.
It was found that including the finite-size effects could delay the onset of the hadron-quark mixed phase and shrink its density range significantly. As the nuclear symmetry energy slope $L$ increases, the onset density of hadron-quark mixed phases decreases. 
From the macroscopic point of view, we see that an increase in the symmetry energy slope $L$ causes a decrease in the maximum mass of hybrid stars by around 0.1 $M_{\odot}$ between $L$ = 30 MeV to 118.5 MeV within the EM method. The results for a hybrid star with a structured mixed phase core are compatible with recent constraints inferred from astrophysical observations.
We also investigated the effects of different surface tensions ($\sigma$), bag constants ($B$), and vector interactions ($G_V$). We found that the pasta phases appear and end earlier for smaller values of $B$ and $G_V$. As $\sigma$ increases, the energy density of pasta phases also increases, leading to a higher onset density and an earlier end density for pasta phases. This results in a smaller radius and mass of the structured mixed phase core.

We investigated the mass and radius of structured mixed phase inside hybrid stars. We observed that, regardless of whether we use the EM method or the GC method, both the radius and mass of the mixed phase in the maximum-mass hybrid star increase with the symmetry energy slope $L$. Compared to the EM method, the GC method tends to have a larger mixed phase core. 
It is noteworthy that, according to the GC method, when the central density of the maximum-mass hybrid star is less than the onset density of quark phase, the hybrid stars may have a mixed phase core rather than a pure quark core. However, within the EM method, the central density of the maximum-mass hybrid star does not monotonically increase with $L$. 
Consequently, whether a pure quark core can occur depends on the model parameters used. Nevertheless, in all cases, we concur that a massive hybrid star ($>$ 2 $M_{\odot}$) may have a structured mixed phase core surrounded by hadronic matter. 
Furthermore, we examined the effects of the parameters $B$ and $G_V$, and observed that smaller values of $B$ and $G_V$ would significantly reduce the maximum mass of hybrid stars.
When the bag constant is set to $B^{1/4}$ = 170 MeV with $G_V$ = 0.2 fm$^2$ and $\sigma$ = 40 MeV fm$^{-2}$, all the cases for $L$ between 30 to 118 MeV support a pure quark core existing in the maximum-mass hybrid star with a radius exceeding 6.0 km.
On the other hand, the radius of the structured mixed phase core with $B^{1/4}$ = 180 MeV and $G_V$ = 0.1 fm$^2$ is comparable to that with $G_V$ = 0.2 fm$^2$, but it has a slightly lower mass.

We examined the effects of the symmetry energy slope $L$ on the initial appearance radius of the pasta phases (droplet, rod, slab, tube, bubble) relative to the center of the maximum-mass hybrid star. For droplet, rod and slab phases, the appearance radius increases with the symmetry energy slope $L$.
Furthermore, we discussed the squared speed of sound and polytropic index of pasta phases inside hybrid stars. The squared speed of sound of the pasta phases remains below 1/3, dropping to zero at the end of the pasta phases. A larger slope $L$ may increases the squared speed of sound of the pasta phases.
The polytropic index $\gamma$ of pasta phases is below 1.75, which is consistent with the criterion that $\gamma <$ 1.75 can be used to distinguish quark degree from hadronic matter.

%%%%%%%%%%%%%%%%%%%%%%%%%%%%%%%%%%%%%%%%%%%%%%%%%%%%%%%%%%%%%%%%%%%%%
\section*{Acknowledgment}
This work was supported in part by the Shandong Natural Science Foundation No. ZR2023QA112; Fundamental Research Funds for the Central Universities under Grant (27RA2210023); National Natural Science Foundation of China No. 12175109 and No. 12305148; Hebei Natural Science Foundation No. A2023203055.

%\bibliography{refs}{}
%\bibliographystyle{aasjournal}
%%%%%%%%%%%%%%%%%%%%%%%%%%%%%%%%%%%%%%%%%%%%%%%%%%%%%%%%%%%%%%%%%%%%%

%%%%%%%%%%%%%%%%%%%%%%%%%%%%%%%%%%%%%%%%%%%%%%%%%%%%%%%%%%%%%%%%%%%%%%%%%%
\end{document}